\documentstyle[12pt,epsf]{article}

\catcode`\@=11
\long\def\@makefntext#1{
\protect\noindent \hbox to 3.2pt {\hskip-.9pt  
$^{{\ninerm\@thefnmark}}$\hfil}#1\hfill}                

\def\@makefnmark{\hbox to 0pt{$^{\@thefnmark}$\hss}}  
        
\def\ps@myheadings{\let\@mkboth\@gobbletwo
\def\@oddhead{\hbox{}
\rightmark\hfil\ninerm\thepage}   
\def\@oddfoot{}\def\@evenhead{\ninerm\thepage\hfil
\leftmark\hbox{}}\def\@evenfoot{}
\def\sectionmark##1{}\def\subsectionmark##1{}}

\renewcommand{\thefootnote}{\fnsymbol{footnote}}

\newcounter{sectionc}\newcounter{subsectionc}\newcounter{subsubsectionc}
\renewcommand{\section}[1] {\vspace*{0.6cm}\addtocounter{sectionc}{1} 
\setcounter{subsectionc}{0}\setcounter{subsubsectionc}{0}\noindent 
        {\normalsize\bf\thesectionc. #1}\par\vspace*{0.4cm}}
\renewcommand{\subsection}[1] {\vspace*{0.6cm}\addtocounter{subsectionc}{1} 
        \setcounter{subsubsectionc}{0}\noindent 
        {\normalsize\it\thesectionc.\thesubsectionc. #1}\par\vspace*{0.4cm}}
\renewcommand{\subsubsection}[1]
{\vspace*{0.6cm}\addtocounter{subsubsectionc}{1}
        \noindent {\normalsize\rm\thesectionc.\thesubsectionc.\thesubsubsectionc. 
        #1}\par\vspace*{0.4cm}}

\newcounter{appendixc}
\newcounter{subappendixc}[appendixc]
\newcounter{subsubappendixc}[subappendixc]

\renewcommand{\appendix}[1] {\vspace*{0.6cm}
        \refstepcounter{appendixc}
        \setcounter{figure}{0}
        \setcounter{table}{0}
        \setcounter{equation}{0}
        \renewcommand{\thefigure}{\Alph{appendixc}.\arabic{figure}}
        \renewcommand{\thetable}{\Alph{appendixc}.\arabic{table}}
        \renewcommand{\theappendixc}{\Alph{appendixc}}
        \renewcommand{\theequation}{\Alph{appendixc}.\arabic{equation}}
        \noindent{\bf Appendix \theappendixc #1}\par\vspace*{0.4cm}}

\def\abstracts#1{{
        \centering{\begin{minipage}{12.2truecm}\footnotesize\baselineskip=12pt\noindent
        \centerline{\footnotesize ABSTRACT}\vspace*{0.3cm}
        \parindent=0pt #1
        \end{minipage}}\par}} 

\renewenvironment{thebibliography}[1]
        {\begin{list}{\arabic{enumi}.}
        {\usecounter{enumi}\setlength{\parsep}{0pt}
\setlength{\leftmargin 1.25cm}{\rightmargin 0pt}
         \setlength{\itemsep}{0pt} \settowidth
        {\labelwidth}{#1.}\sloppy}}{\end{list}}

\newcounter{itemlistc}
\newcounter{romanlistc}
\newcounter{alphlistc}
\newcounter{arabiclistc}

\newcommand{\fcaption}[1]{
        \refstepcounter{figure}
        \setbox\@tempboxa = \hbox{\footnotesize Fig.~\thefigure. #1}
        \ifdim \wd\@tempboxa > 6in
           {\begin{center}
        \parbox{6in}{\footnotesize\baselineskip=12pt Fig.~\thefigure. #1}
            \end{center}}
        \else
             {\begin{center}
             {\footnotesize Fig.~\thefigure. #1}
              \end{center}}
        \fi}

\newcommand{\tcaption}[1]{
        \refstepcounter{table}
        \setbox\@tempboxa = \hbox{\footnotesize Table~\thetable. #1}
        \ifdim \wd\@tempboxa > 6in
           {\begin{center}
        \parbox{6in}{\footnotesize\baselineskip=12pt Table~\thetable. #1}
            \end{center}}
        \else
             {\begin{center}
             {\footnotesize Table~\thetable. #1}
              \end{center}}
        \fi}

\def\@citex[#1]#2{\if@filesw\immediate\write\@auxout
        {\string\citation{#2}}\fi
\def\@citea{}\@cite{\@for\@citeb:=#2\do
        {\@citea\def\@citea{,}\@ifundefined
        {b@\@citeb}{{\bf ?}\@warning
        {Citation `\@citeb' on page \thepage \space undefined}}
        {\csname b@\@citeb\endcsname}}}{#1}}

\newif\if@cghi
\def\cite{\@cghitrue\@ifnextchar [{\@tempswatrue
        \@citex}{\@tempswafalse\@citex[]}}
\def\citelow{\@cghifalse\@ifnextchar [{\@tempswatrue
        \@citex}{\@tempswafalse\@citex[]}}
\def\@cite#1#2{{$\null^{#1}$\if@tempswa\typeout
        {IJCGA warning: optional citation argument 
        ignored: `#2'} \fi}}

 1
 1
 1

\font\ninerm=cmr9

\begin{document}

{\hfill MPI/PhT/96--130}\\
\mbox{}
{\hfill hep-ph/9612395}\\
\mbox{}
{\hfill December 1996}
\vspace*{0.5cm}

\centerline{\normalsize\bf 
  CORRECTIONS OF ${\cal O}(\alpha_s^2)$ TO THE DECAY OF AN
  INTERMEDIATE--MASS
}
\baselineskip=22pt
\centerline{\normalsize\bf 
  HIGGS BOSON INTO TWO 
PHOTONS\footnote{Contribution to the proceedings of the Ringberg Workshop, 
          December 1996.}
}
\baselineskip=22pt

\centerline{\footnotesize MATTHIAS STEINHAUSER}
\baselineskip=13pt
\centerline{\footnotesize\it 
Max-Planck-Institut f\"ur Physik (Werner-Heisenberg-Institut),
}
\centerline{\footnotesize\it 
F\"ohringer Ring 6, 80805 Munich, Germany
}
\baselineskip=22pt

\vspace*{0.9cm}
\abstracts{
The QCD correction of ${\cal O}(\alpha_s^2)$ to the decay of the
Standard Model Higgs boson into two photons is presented.
We consider the contribution coming from diagrams with 
a heavy top quark as
virtual particle. The first three terms of the expansion
in the inverse top mass is calculated.
Expressing the result through the on-shell top mass $M_t$, we find
large coefficients for the power-suppressed terms whereas in the
$\overline{\mbox{MS}}$ scheme the coefficients are tiny. 
}
 
\vspace*{0.6cm}
\normalsize\baselineskip=15pt
\setcounter{footnote}{0}
\renewcommand{\thefootnote}{\alph{footnote}}

The Higgs boson is the only undetected particle in the Standard Model (SM).
Now that top quark is well established 
\cite{CDFD0}
the discovery of the Higgs
boson is one of the most important aims in high-energy physics.
Current experiments at LEP ruled out a Higgs boson
with mass $M_H\le65.6$~GeV
\cite{Jan96}
via Bjorken's process $e^+e^-\to Z\to f\bar{f}H$.
The next generation of colliders will improve this limit.
Despite the fact that 
for an intermediate-mass Higgs boson with $M_W\le M_H\le 2M_W$
the decay into a b quark pair will be the dominant
process, a detection in this channel is
unlikely with the LHC because of the large background.
A very promising channel for the discovery 
especially for $M_H<130$~GeV 
is the decay into two
photons, though the branching ratio is of ${\cal O}(10^{-3})$.
$\Gamma(H\to\gamma\gamma)$ also enters the cross section for
Higgs-boson production through photon-photon fusion, which is the
dominant process in future high energy $e^+e^-$ colliders.
Therefore it is important to calculate higher order corrections to
this process.
In this note we consider the ${\cal O}(\alpha_s^2)$ corrections
to $\Gamma(H\to\gamma\gamma)$ coming from a top quark which is
heavy compared to the Higgs boson.
Besides the leading term which is independent of the top-quark mass,
we also calculate the first two power-suppressed corrections.
Throughout this paper we work in the framework of dimensional regularisation
with $D=4-2\epsilon$. Large intermediate terms are treated with
the help of FORM
\cite{Ver91}.

The coupling $H\gamma\gamma$ does not exist at Born level, so
the decay of the Higgs boson into photons is a loop-induced process.
At the one-loop level either fermions or $W$ bosons may be present
in the loop, and the decay rate reads
\begin{equation}
\Gamma(H\to\gamma\gamma) = 
     \Big| \sum_{f} A_f(\tau_f) + A_W(\tau_W) \Big|^2\frac{M_H^3}{64\pi},
\end{equation}
with $\tau_f=M_H^2/4m_f^2$ and $\tau_W=M_H^2/4M_W^2$.
In the one-loop order $A_W$ is dominating the correction.
The contribution from light fermions is suppressed by their mass.
We will focus our attention to $A_t(\tau_t)$ and consider
QCD corrections of ${\cal O}(\alpha_s^2)$ to this amplitude. 
It is convenient to write
\begin{equation}
A_t(\tau_t) = A_t^{(0)} + \frac{\alpha_s}{\pi} A_t^{(1)}
            + \left(\frac{\alpha_s}{\pi}\right)^2 A_t^{(2)}
            + \ldots\,,
\end{equation}
where $\alpha_s\equiv\alpha_s^{(6)}(\mu)$ throughout this paper.

There is
no decoupling in the limit $M_t\to\infty$ because the 
$H\bar{t}t$ coupling is proportional to the top quark mass.
The exact one-loop result is well-known and reads for $\tau_t\le 1$
\cite{EllGaiNan76,VaiVolSakShi79}
\begin{eqnarray}
A_t^{(0)} &=& \hat{A}_t\left[
     \frac{3}{2\tau_t}\left(
                1+\left(1-\frac{1}{\tau_t}
                     \right)\arcsin^2\sqrt{\tau_t}\right)
                       \right]
\nonumber\\
          &=& \hat{A}_t\left(1
                            +\frac{7}{30}\tau_t
                            +\frac{2}{21}\tau_t^2
                            +\frac{26}{525}\tau_t^3
                            +\frac{512}{17325}\tau_t^4
                            +\frac{1216}{63063}\tau_t^5
                            +{\cal O}(\tau_t^6)
                       \right) 
\label{oneloop}
\end{eqnarray}
with $\hat{A}_t=N_C\frac{2\alpha}{3\pi v}Q_t^2$ and $v=2^{-1/4}G_F^{-1/2}$. 
The second line of Eq.~(\ref{oneloop}) contains
the expansion for large $M_t$.

\begin{figure}[ht]
 \begin{center}
  \begin{tabular}{c}
    \epsfxsize=12.0cm
    \leavevmode
    \epsffile[50 150 500 650]{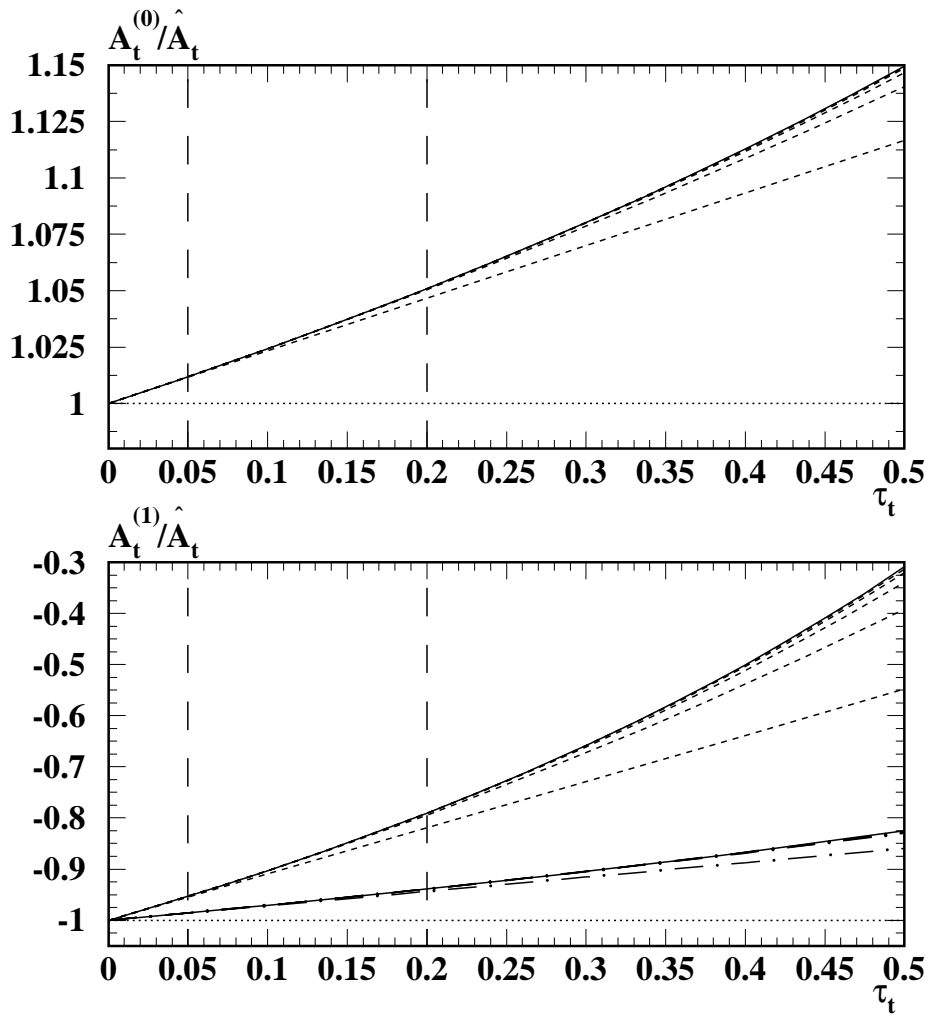}
  \end{tabular}
  \caption{\label{fig12}One- and two-loop corrections to the amplitude $A_t$.
           Successively higher orders in $\tau_t$ are included.
           Dotted line: leading order result,
           dashed line: on-shell scheme; 
           dash-dotted line: $\overline{\mbox{MS}}$ scheme, full line:
           exact result. The two vertical lines indicate $M_H=M_W$ and
           $M_H=2M_W$, respectively.}
 \end{center}
\end{figure}

At the two-loop level also the QCD corrections are at hand.
Whereas for arbitrary values of $M_H$ and $M_t$ only a numerical
result is available 
\cite{DjoSpiVdbZer91} 
in the limit of a heavy top quark also analytical results exist
\cite{DawKau93}.
We repeated this calculation 
and even extended it
to get the first six terms in the expansion for a heavy
quark. The result is given by 
\begin{eqnarray}
A_t^{(1)} &=& \hat{A}_t 
              \left(-1 + \frac{122}{135}\tau_t
                       + \frac{8864}{14175}\tau_t^2
                       + \frac{209186}{496125}\tau_t^3
                       + \frac{696616}{2338875}\tau_t^4
\right.
\nonumber\\
&&
\left.
\hphantom{\hat{A}_t \frac{\alpha_s}{\pi}}
                       + \frac{54072928796}{245827456875}\tau_t^5
                       + {\cal O}(\tau_t^6)
              \right).
\end{eqnarray}
In order to get an impression of the quality of the expansion for 
large $M_t$ we plot in Fig.~\ref{fig12}
the one- and two-loop results for $A_t$ against $\tau_t$.
The dotted lines are the result for infinitely large top mass. 
In the other curves successively higher orders of $\tau_t$
are included. The two vertical lines indicate $M_H=M_W$ and 
$M_H=2 M_W$, respectively. Also the exact results are shown
(full line). 
At two loops the result in the on-shell scheme (dashed line)
differs from the one in the $\overline{\mbox{MS}}$ scheme
(dash-dotted line, $\mu^2=\bar{m}_t^2$; here the variable of the
abscissa is $\bar{\tau}_t(\mu)=M_H^2/4\bar{m}_t^2(\mu)$). 
This numerical difference is compensated 
by an appropriate change of the one-loop result. 
It should be noted that in this context the transitition to the
$\overline{\mbox{MS}}$ scheme is done by replacing the on-shell
top mass $M_t$ through the running mass $\bar{m}_t(\mu)$.

For $M_H=\sqrt{2}M_t\approx 250$~GeV (i.e. $\tau_t=0.5$)
the curves including the $\tau_t^5$ term  are practically
indistinguishable from the exact result.
But already the curves which only contain the 
$\tau_t^2$ term deliver a reasonable result and
for $M_H\approx 2 M_W$ this correction 
is a very good approximation.
For $M_H\approx M_W$ the inclusion of the linear term only
is enough to describe the exact result very well. 
We take these
considerations as a motivation to calculate in the 
three-loop case the terms proportional to $\tau_t^2$.

There are different methods which allow one to calculate of 
$\Gamma(H\to \gamma\gamma)$ in the limit that a heavy virtual quark 
is present in the loop and the 
Higgs boson is in the intermediate mass range
\cite{Stediss}.

A very elegant method to get the leading order
(i.e. the limit $M_t\to \infty$) result is provided by the use of
a low-energy theorem (LET) for the Higgs boson
\cite{EllGaiNan76,VaiVolSakShi79,KniSpi95}.
The basic idea is that the coupling of the Higgs boson to 
other particles is proportional to the mass of these particles.
Both for fermions and gauge bosons the substitution $m_i\to m_i(1+H/v)$
generates the coupling of the particle with mass $m_i$ to the Higgs boson.
For a Higgs boson carrying zero four-momentum
this is obtained by differentiating the respective two-point 
function with respect to the virtual-particle masses. To be 
more precise, in our case we can write down the effective Lagrangian
\begin{equation}
{\cal L}_{\gamma\gamma} = -\frac{1}{4} F^{0,\mu\nu}F^{0}_{\mu\nu}
               \left(1+\tilde{\Pi}_{\gamma\gamma}^{0,t}(0)\right)
\end{equation}
where $F^{0}_{\mu\nu}$ is the field strength tensor and 
$\tilde{\Pi}_{\gamma\gamma}^{0,t}(0)=
        \Pi_{\gamma\gamma}^{0,t}(q^2)/q^2|_{q=0}$.
$\Pi_{\gamma\gamma}^{0,t}(q^2)$
is (the transversal part of) the photon 
two-point function containing the top quark as virtual particle.
The superscript zero indicates that we are still dealing with
bare quantities.
Applying the LET we get an effective Lagrangian for the $H\gamma\gamma$
vertex
\begin{equation}
{\cal L}_{H\gamma\gamma} = -\frac{1}{4} F^{0,\mu\nu}F^{0}_{\mu\nu}
                          \frac{m_t^0\partial}{\partial m_t^0}
              \tilde{\Pi}_{\gamma\gamma}^{0,t}(0)
              \frac{H}{v}.
\label{laghgg}
\end{equation}
It is clear that after the differentiation the renormalization
of the coupling constant $\alpha_s^0$ and of $m_t^0$ has to be performed.

The method of the LET has been used very extensively in the past. Besides 
effective $H\gamma\gamma$ and $Hgg$ couplings also the $H\bar{b}b$, 
$HWW$ and $HZZ$
couplings where considered at the 
two-
\cite{KniSpi94,KniSpi95}
and even three-loop levels
\cite{KniSte95,CheKniSte96}.
In a recent work the theoretical background of the LET was considered
\cite{Kil95}.
The big advantage of the LET is that much less diagrams have to
be taken into account and the three-point amplitudes are simply
generated by differentiating w.r.t the top-quark mass.

The result for 
$\tilde{\Pi}_{\gamma\gamma}^{0,t}(0)$
is given by
\cite{CheKueSte96}
\begin{eqnarray}
\tilde{\Pi}_{\gamma\gamma}^{0,t}(0) &=& N_C Q_t^2\frac{\alpha}{4\pi}
\Bigg\{ \frac{4}{3\epsilon} + \frac{4}{3} L_0 
       + \frac{\alpha_s}{\pi}C_F \left(\frac{13}{12} - \frac{3}{2\epsilon} 
                                     - 3 L_0 \right)
\nonumber\\
&&
       + \left(\frac{\alpha_s}{\pi}\right)^2  \Bigg[
        C_F  \left(  - \frac{7}{16} + \frac{7}{32} \zeta(3) \right)
\nonumber\\
&&
        +C_A C_F \left(  - \frac{4243}{1296} 
                         - \frac{11}{18\epsilon^2}
                         - \frac{11}{6\epsilon} L_0 
                         - \frac{5}{108\epsilon}
                         + \frac{223}{96} \zeta(3) 
                         - \frac{11}{12} 
                         - \frac{5}{36} L_0 
                         - \frac{11}{4} L_0^2
                  \right)
\nonumber\\
&&
       + C_F^2  \left(  - \frac{137}{72} 
                        + \frac{17}{6\epsilon} 
                        - \frac{95}{48} \zeta(3) 
                        + \frac{17}{2} L_0 
                \right)
\nonumber\\
&&
       + C_F T n_f\left( \frac{169}{162} 
                     + \frac{2}{9\epsilon^2} 
                     + \frac{2}{3\epsilon} L_0 
                     - \frac{5}{27\epsilon}
                     + \frac{1}{3}\zeta(2) 
                     - \frac{5}{9}L_0 
                     + L_0^2 
                \right)
\Bigg]
\Bigg\}
\end{eqnarray}
where $N_C=3$, $C_F=4/3$, $C_A=3$, $L_0=\ln(\mu^2/(M_t^0)^2)$, $T=1/2$
and $n_f=6$ 
is the total number of quarks.
If this expression is inserted into Eq.~(\ref{laghgg})
and the top-mass and coupling-constant renormalization is performed 
we receive the
leading-order term for the amplitude of the Higgs-boson decay into a
pair of photons
\begin{eqnarray}
A_t^{\mbox{\scriptsize lo}} &=& \hat{A}_t
\Bigg\{ 1 
             - \frac{\alpha_s}{\pi} C_F \frac{3}{4}
\label{leadord}
\\&&
             +\left(\frac{\alpha_s}{\pi}\right)^2
\Bigg[
         C_F C_A  \left(  - \frac{7}{12} - \frac{11}{16} L \right)
       + C_F^2              \frac{27}{32}
       + C_F T n_f  \left(  - \frac{1}{12} + \frac{1}{4} L  \right)
       - C_F T            \frac{3}{16} 
\nonumber
\Bigg]
\Bigg\}
\end{eqnarray}
with $L=\ln(\mu^2/M_t^2)$.
In this paper the diagrams with the Higgs boson and the 
photons attached to different fermion lines are not
considered. They will be treated elsewhere
\cite{hphoton2}.
For this reason in order to compare the result from the LET
with the explicit calculation, the contributions from the
double-bubble diagram coming from the differentiation w.r.t. the 
inner top mass have to be subtracted. This
is already done in (\ref{leadord}).

\begin{figure}[ht]
 \leavevmode
 \begin{center}
 \begin{tabular}{cccc}
   \epsfxsize=2.8cm
   \epsffile[213 232 399 484]{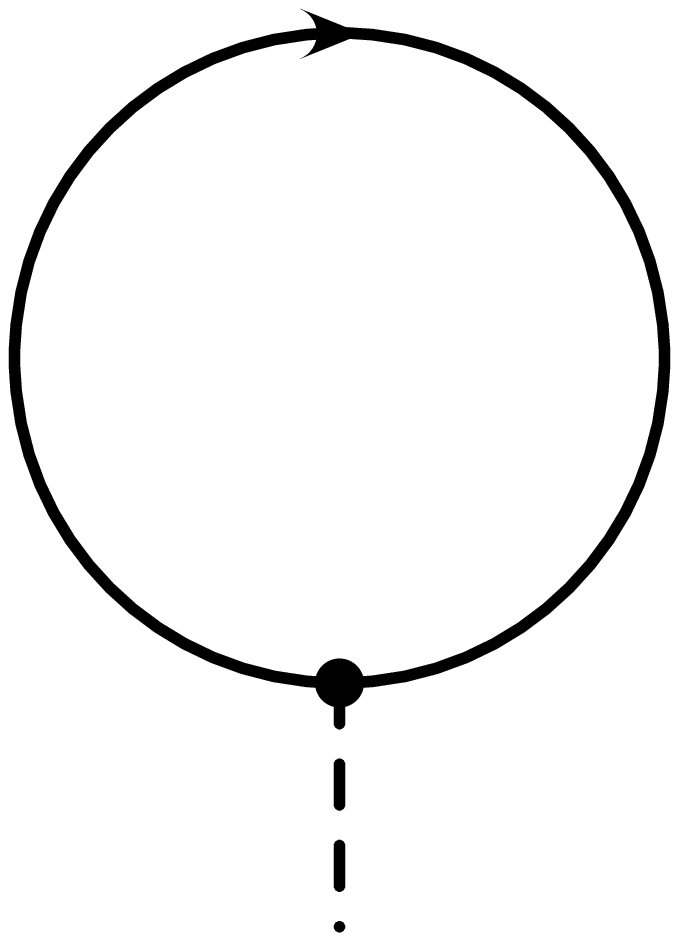}
&
   \epsfxsize=2.8cm
   \epsffile[213 232 399 484]{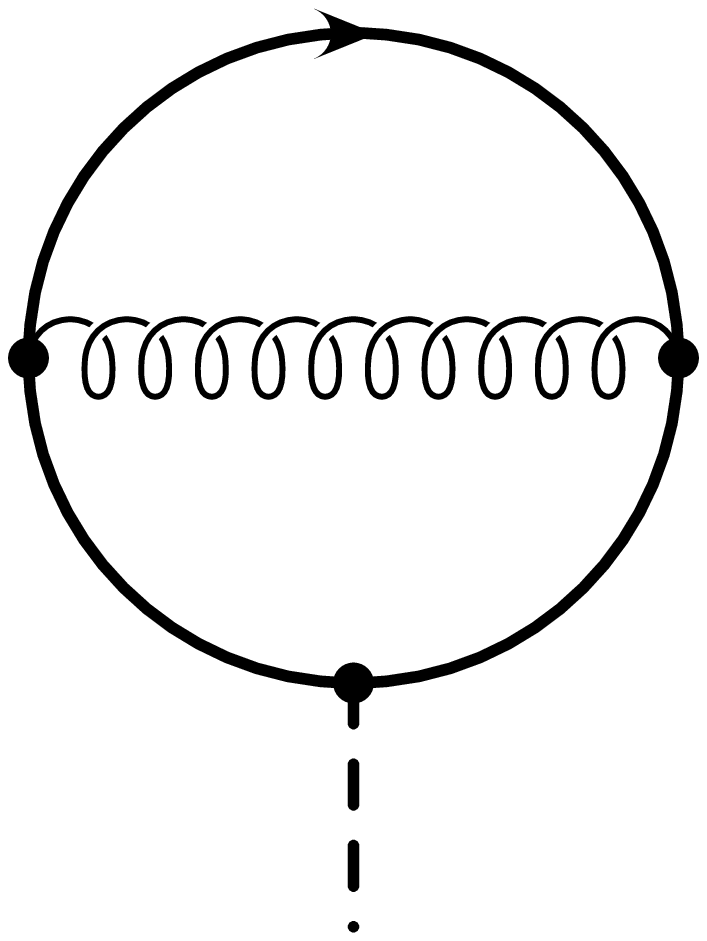}
&
   \epsfxsize=2.8cm
   \epsffile[213 232 399 484]{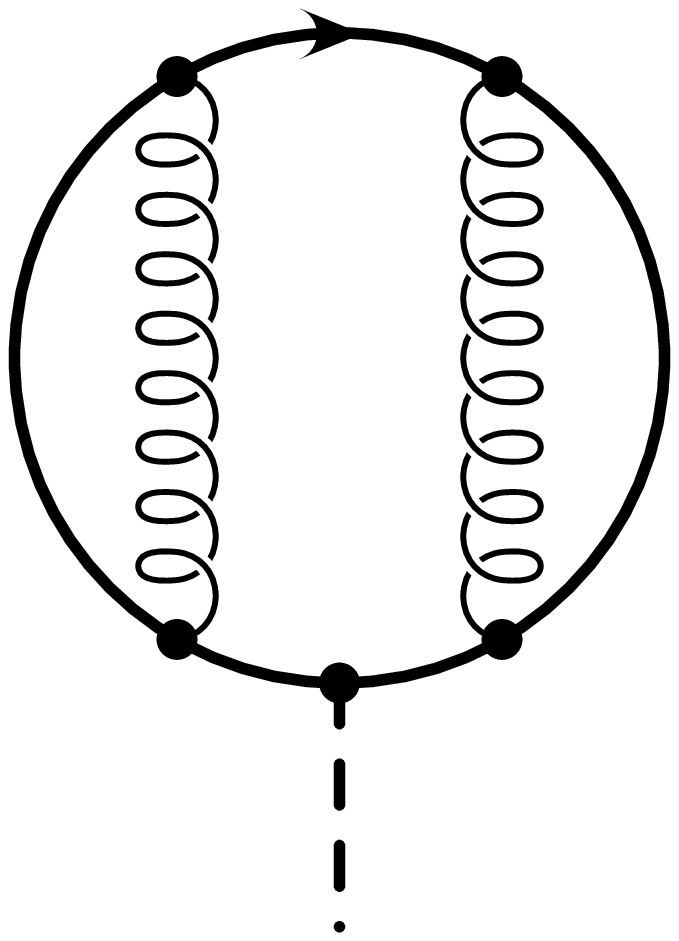}
&
   \epsfxsize=2.8cm
   \epsffile[213 232 399 484]{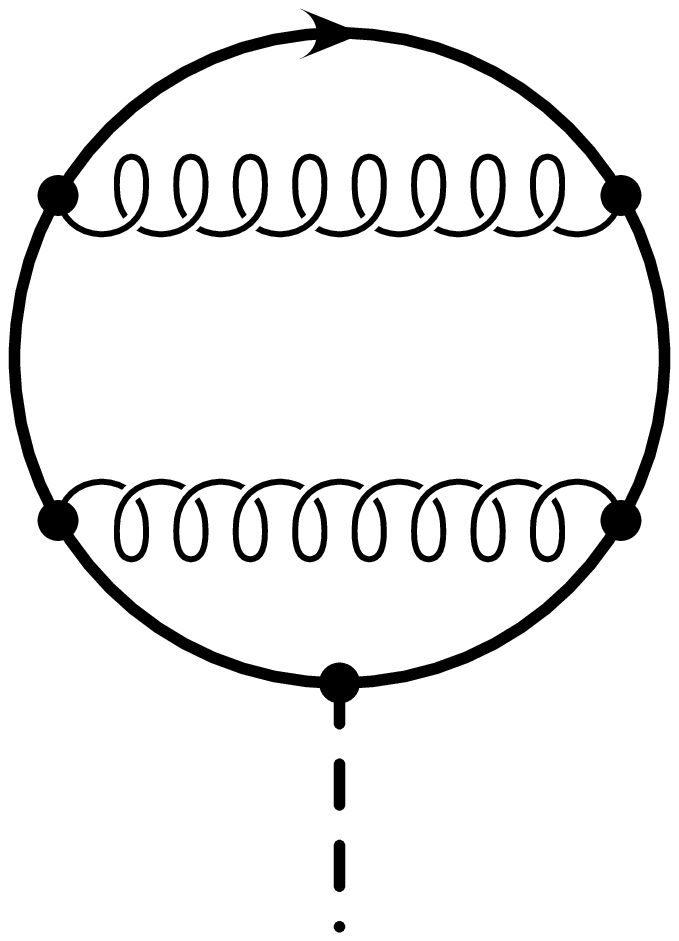}
\\
\\
   \epsfxsize=2.8cm
   \epsffile[213 232 399 484]{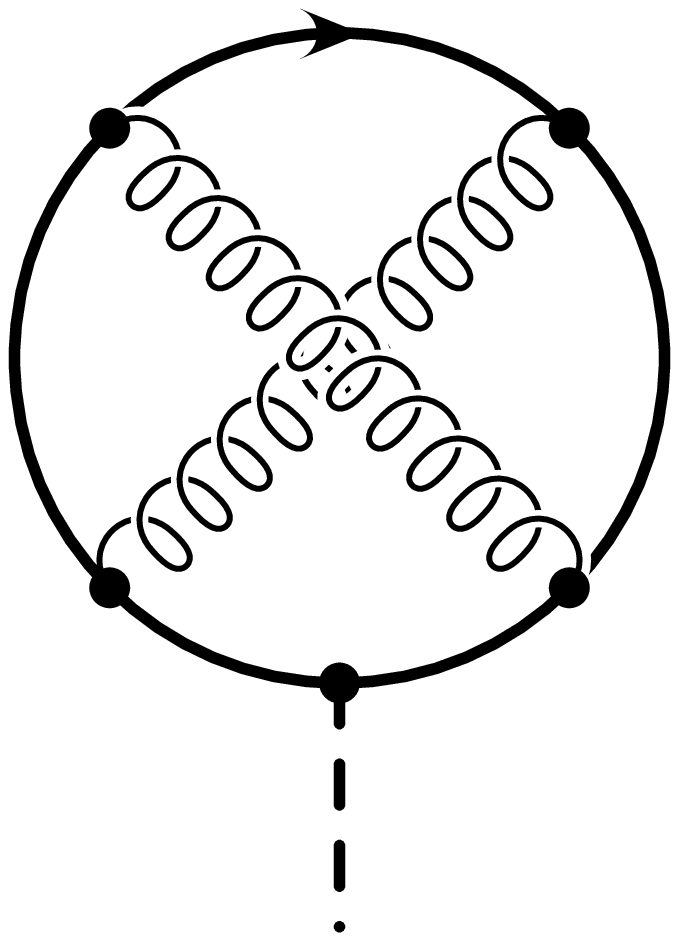}
&
   \epsfxsize=2.8cm
   \epsffile[213 232 399 484]{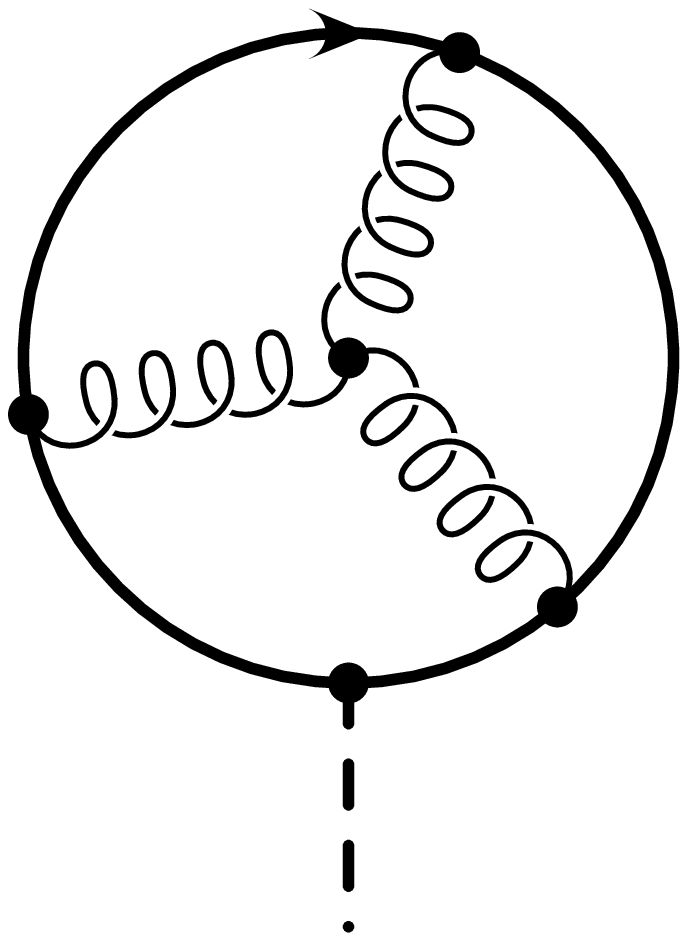}
&
   \epsfxsize=2.8cm
   \epsffile[213 232 399 484]{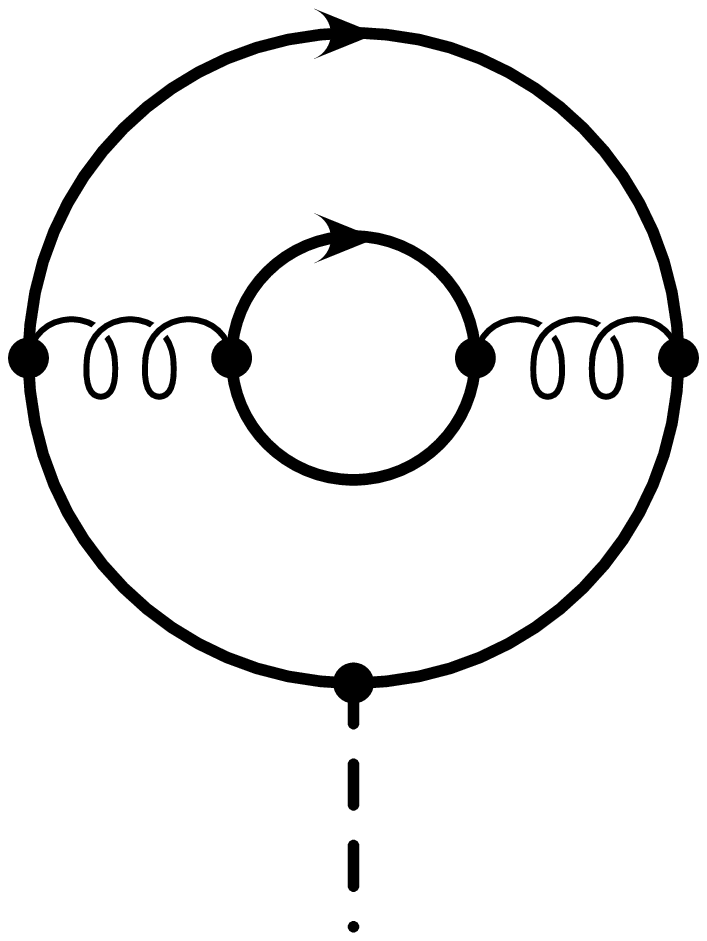}
&
   \epsfxsize=2.8cm
   \epsffile[213 232 399 484]{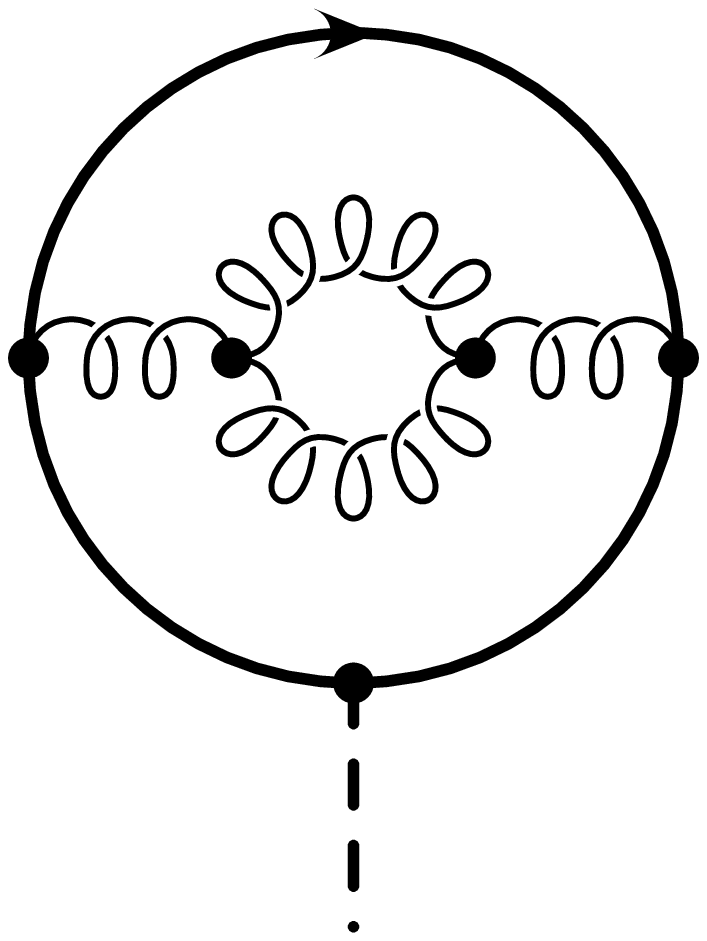}
 \end{tabular}
 \caption{\label{figfsg}
          Tadpole diagrams relevant for the method based on the 
          Fock-Schwinger gauge. The diagram with a second light
          quark loop and the one containing the ghost particle are not
          displayed.}
 \end{center}
\end{figure}

A second method to compute the leading-order result is provided
by the use of the so-called Fock-Schwinger gauge for the external
photons. In coordinate space this reads as $x^\mu A_\mu(x)=0$.
It can be shown
\cite{NovShiVaiZak84}
that from this condition the expansion of $A_\mu(x)$ in terms of local 
operators has the form
\begin{eqnarray}
A_\mu(x) &=& \frac{1}{2\cdot0!} x^\rho F_{\rho\mu}(0) 
           + \frac{1}{3\cdot1!} x^\sigma x^\rho 
                                     \left( D_\sigma F_{\rho\mu}(0) \right)
           + \ldots\,\,.
\label{eqA}
\end{eqnarray}
In the present case we are interested in attaching two photons
to the diagrams pictured in Fig.~\ref{figfsg}. If only the leading order
is considered it is enough to keep the first term
in Eq.~(\ref{eqA}). Then the prescription derived in
\cite{NovShiVaiZak84}
is as follows:
{\it (i)} Route the momentum associated with the photon
through the diagram to the Higgs boson; 
{\it (ii)} differentiate the affected propagators w.r.t. this momentum;
{\it (iii)} set the momentum equal to zero
and {\it (iv)} proceed in the same way with the second photon.
At the three-loop level there are eight tadpole diagrams which have to be 
considered. A small computer program written in FORM
\cite{Ver91}
applies the above algorithm and attaches two photons.
Afterwards the integration is performed.
With this method the same result as in Eq.~(\ref{leadord})
is obtained. 

There is also a method which leads directly to the decay rate.
Thereby one has to consider $\Pi^H(q^2)$,
the two point function for the Higgs boson.
The imaginary part of $\Pi^H(q^2)$ coming from a two photon cut
has to be extracted. In order to get the width the relation
\begin{equation}
\Gamma(H\to\gamma\gamma)=\frac{1}{M_H}\mbox{Im}\Pi^H(M_H^2)
\end{equation}
can be used.
In our case, it is convenient to consider double-triangle diagrams where
in both triangles top quarks (and possibly gluons) are present.
In the limit that the external momentum $q^2=M_H^2$ is less than 
$4M_t^2$, the only imaginary part comes from the two-photon cut.
With this procedure we where able to reproduce the results of Ref.
\cite{DawKau93}.

The most obvious method is the expansion of the triangle diagrams
in their external momenta $q_1$ and $q_2$. One thus successively
obtains higher-order terms in $M_H^2/M_t^2$ depending on how far
the expansion is performed.
The amplitude for the decay of the Higgs boson into two photons
with polarization vectors $\epsilon_\mu(q_1)$ and $\epsilon_\nu(q_2)$
has the following Lorentz structure:
\begin{eqnarray}
A_t^{\mu\nu}&=&\sum_{i} A_{t,i}^{\mu\nu} 
          =\sum_{i} \left( a_{t,i}\, q_1q_2\, g^{\mu\nu}
                         + b_{t,i}\, q_1^{\nu}q_2^{\mu}
                         + c_{t,i}\, q_1^{\mu}q_2^{\nu} \right),
\label{amunu}
\end{eqnarray}
where $c_{t,i}$ has no contribution for on-shell photons.
In Eq.~(\ref{amunu}) the sum runs over all diagrams relevant for
the decay $H\to\gamma\gamma$.
Due to gauge invariance, we have $\sum_i a_{t,i} = -\sum_i b_{t,i}$.
It is easy to find projectors for $a_{t,i}$ and $b_{t,i}$:
\begin{eqnarray}
a_{t,i} &=& \frac{ A_{t,i}^{\mu\nu} }{(D-2)(q_1q_2)^2} 
  \left(q_1q_2\, g_{\mu\nu} - q_{1\nu}q_{2\mu} - q_{1\mu}q_{2\nu}\right),\\
b_{t,i} &=& \frac{ A_{t,i}^{\mu\nu} }{(D-2)(q_1q_2)^2} 
  \left(-q_1q_2\, g_{\mu\nu} + q_{1\nu}q_{2\mu} + (D-1)q_{1\mu}q_{2\nu}\right).
\end{eqnarray}
In fact, we calculated both, $a_{t,i}$ and $b_{t,i}$, 
in order to have an additional
check for the correctness of our result.
A ($2D$ dimensional) Taylor expansion of the 
triangle diagrams in the two external momenta $q_1$ and $q_2$ 
leads to the following structure for $a_{t,i}(q_1,q_2)$:
\begin{equation}
a_{t,i}(q_1,q_2)=\sum_{l,m,n=0}^{\infty} c_{lmn}^{i}
                          (q_1^2)^l (q_2^2)^m (q_1q_2)^n.
\end{equation}
After projecting out the $c_{00n}^{i}$, 
we end up with massive 
three-loop diagrams with zero external momentum. 
For these integrals the technique is meanwhile well established
\cite{CheKueSte96,drho}.
After the decomposition of the numerator in terms of the denominator,
the scalar integrals are reduced via recurrence relations
\cite{CheTka81Bro92}
to trivial ones or so-called master integrals, for
which a hard three-loop calculation is necessary.

\begin{figure}[ht]
 \begin{center}
    \epsfxsize=4.0cm
    \leavevmode
    \epsffile[189 314 435 478]{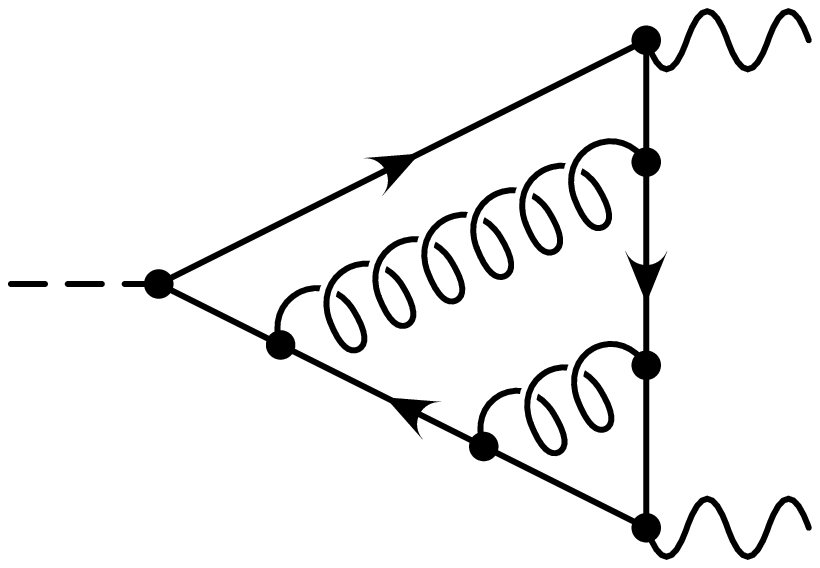}
    \epsfxsize=4.0cm
    \leavevmode
    \epsffile[189 314 435 478]{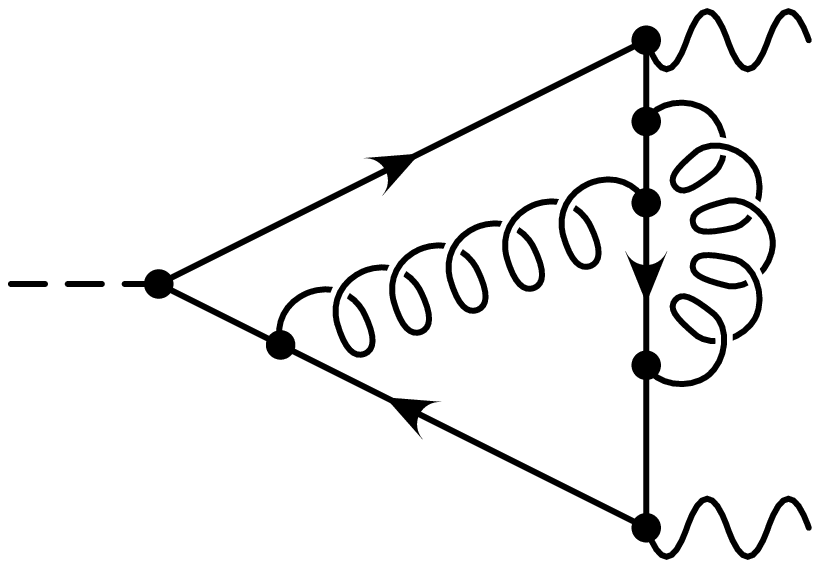}
    \epsfxsize=4.0cm
    \leavevmode
    \epsffile[189 314 435 478]{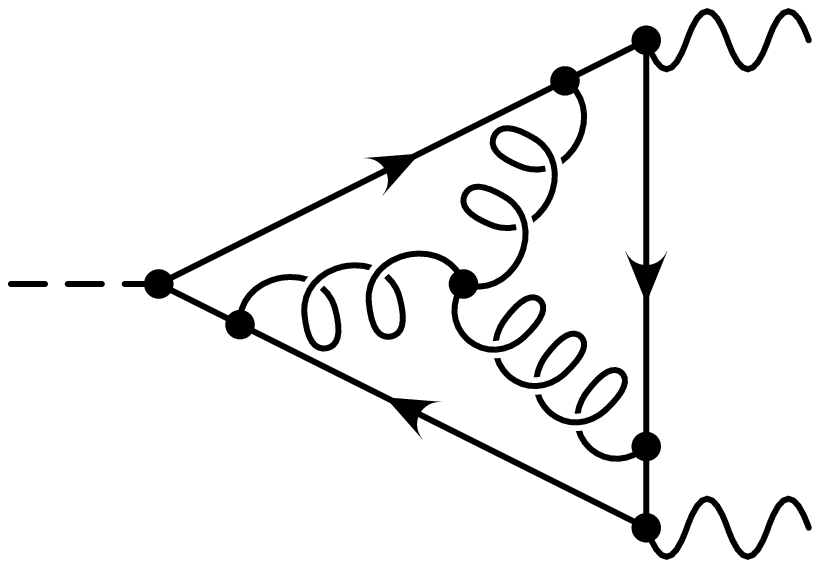}
  \caption{ \label{figdia} Typical Feynman diagrams contributing
                           to $H\to\gamma\gamma$. }
 \end{center}
\end{figure}

\begin{figure}[ht]
 \begin{center}
  \begin{tabular}{c}
    \epsfxsize=12.0cm
    \leavevmode
    \epsffile[60 300 500 530]{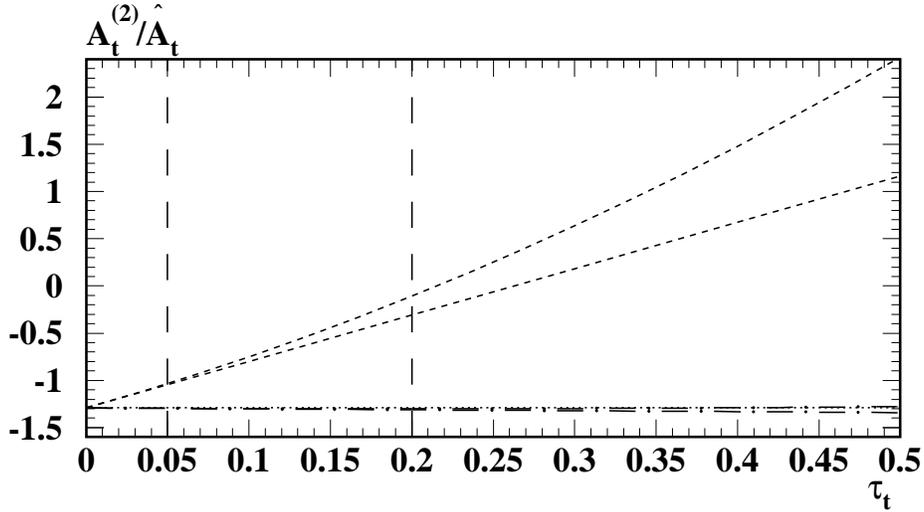}
  \end{tabular}
  \caption{\label{fig3}${\cal O}(\alpha_s^2)$ corrections to 
           $H\to\gamma\gamma$.
           The same notation is adopted as in Fig. \ref{fig12}. }
 \end{center}
\end{figure}

At ${\cal O}(\alpha_s^2)$, altogether $2\times79$ diagrams (the ``2''
comes from the two possible orientations of the fermion line)
have to be taken into account.
In Fig.~\ref{figdia}, three exemplary graphs are pictured.
Setting $C_A=3, C_F=4/3$ and $n_f=6$, the result reads
($A_t = \sum_i a_{t,i}$)
\cite{Stediss}:
\begin{eqnarray}
A_t &=& 
     \hat{A}_t \Bigg[1  + \frac{7}{30}\tau_t + \frac{2}{21} \tau_t^2
                            +\frac{26}{525}\tau_t^3
                            +\frac{512}{17325}\tau_t^4
                            +\frac{1216}{63063}\tau_t^5 
\nonumber\\
&& 
        + \frac{\alpha_s}{\pi} 
           \left(-1 + \frac{122}{135} \tau_t 
                    + \frac{8864}{14175} \tau_t^2
                       + \frac{209186}{496125}\tau_t^3
                       + \frac{696616}{2338875}\tau_t^4
                       + \frac{54072928796}{245827456875}\tau_t^5
           \right)
\nonumber\\
&& 
         +\left(\frac{\alpha_s}{\pi}\right)^2
          \left(-\frac{31}{24} - \frac{7}{4} \ln\frac{\mu^2}{M_t^2} 
\right.
\nonumber\\
&& 
\left.
         +\left(-\frac{19531913}{622080}
               +\frac{7}{45}\,\zeta(2)
               +\frac{14}{45}\,\zeta(2)\ln(2) 
               +\frac{821063}{27648}\,\zeta(3)
               +\frac{427}{270} \ln\frac{\mu^2}{M_t^2}\right) \tau_t
\right.
\nonumber\\
&&
\left.
         +\left(-\frac{56709666623}{2612736000}
                +\frac{8}{63}\,\zeta(2)
                +\frac{16}{63}\,\zeta(2)\ln(2)
                +\frac{72438107}{3317760}\,\zeta(3)
\right.
\right.
\nonumber\\
&&
\left.
\left.
           + \frac{2216}{2025} \ln\frac{\mu^2}{M_t^2}\right) \tau_t^2\right)
         +\ldots
         \Bigg]
\nonumber\\
    &=& \hat{A}_t \Bigg[1  + 0.2333 \tau_t + 0.09524 \tau_t^2
                           + 0.04952\tau_t^3+ 0.02955\tau_t^4
                           + 0.01928\tau_t^5  
\nonumber\\
&&
         + \frac{\alpha_s}{\pi} \left(-1 + 0.9037 \tau_t 
                                     + 0.6253 \tau_t^2
                                     + 0.4216 \tau_t^3   
                                     + 0.2978 \tau_t^4  
                                     + 0.2200 \tau_t^5 
                                \right)
\nonumber\\
&& 
         +\left(\frac{\alpha_s}{\pi}\right)^2
          \left(-1.292 - 1.75 \ln\frac{\mu^2}{M_t^2} 
         +\left(4.91035 + 1.58148 \ln\frac{\mu^2}{M_t^2}\right) \tau_t\right.
\nonumber\\
&&
\hphantom{\left(\frac{\alpha_s}{\pi}\right)}
         +\left.
         \left(5.038 + 1.094 \ln\frac{\mu^2}{M_t^2}\right) \tau_t^2\right)
         +\ldots
         \Bigg],
\label{resos}
\end{eqnarray}
where the dots represent higher orders in $\tau_t$ and $\alpha_s$.
The leading-order result of this method coincides with Eq.~(\ref{leadord}).
To the knowledge of the author, this is the first time that the
LET was tested at three-loop level.
In Eq.~(\ref{resos}),
the on-shell definition of the pole mass was used. Expressing the
result in terms of the $\overline{\mbox{MS}}$ mass
\cite{GraBroGraSch90}, 
$\bar{m}_t(\mu)$, the
result looks like:
\begin{eqnarray}
\bar{A}_t 
    &=& \hat{A}_t \Bigg[1  + \frac{7}{30} \bar{\tau}_t 
                           + \frac{2}{21} \bar{\tau_t}^2
                            +\frac{26}{525}\bar{\tau_t}^3
                            +\frac{512}{17325}\bar{\tau_t}^4
                            +\frac{1216}{63063}\bar{\tau_t}^5 
\nonumber\\
&&
     + \frac{\alpha_s}{\pi} \left(-1 
         + \left(\frac{38}{135} - \frac{7}{15}\bar{l}\right) \bar{\tau}_t
         + \left(\frac{1664}{14175} - \frac{8}{21} \bar{l}\right)\bar{\tau}_t^2
         + \left(\frac{12626}{496125} 
                 - \frac{52}{175} \bar{l}\right)\bar{\tau}_t^3
\right.
\nonumber\\
&&
\left.
\hphantom{\left(\frac{\alpha_s}{\pi}\right)}
         + \left(-\frac{40664}{2338875} 
                    - \frac{4096}{17325} \bar{l}\right)\bar{\tau}_t^4
         + \left(-\frac{9128671204}{245827456875} 
                    - \frac{12160}{63063} \bar{l}\right)\bar{\tau}_t^5
\right)
\nonumber\\
&& 
         +\left(\frac{\alpha_s}{\pi}\right)^2
          \left(-\frac{31}{24} - \frac{7}{4} \bar{l}
         +\left(-\frac{22326329}{622080}
                +\frac{4116067}{138240}\,\zeta(3) 
                - \frac{769}{1080}\bar{l} 
                + \frac{7}{120} {\bar{l}}\,^2 \right) \bar{\tau}_t\right.
\nonumber\\
&& 
\hphantom{\left(\frac{\alpha_s}{\pi}\right)}
          \left.
         +\left(-\frac{68094821183}{2612736000}
               +\frac{508541309}{23224320}\,\zeta(3)
               -\frac{1241}{1575} \bar{l} 
               +\frac{3}{7}{\bar{l}}\,^2 \right) \bar{\tau}_t^2\right)
         +\ldots
         \Bigg]
\nonumber\\
    &=& \hat{A}_t \Bigg[1  + 0.2333 \bar{\tau}_t  + 0.09524 \bar{\tau}_t^2
                           + 0.04952\bar{\tau}_t^3+ 0.02955 \bar{\tau}_t^4
                           + 0.01928\bar{\tau}_t^5  
\nonumber\\
&&
     + \frac{\alpha_s}{\pi} \left(-1 + (0.2815 
                           - 0.4667\bar{l}) \bar{\tau}_t
                  + (0.1174 - 0.3810 \bar{l})\bar{\tau}_t^2
                  + (0.02545 - 0.2971 \bar{l})\bar{\tau}_t^3
\right.
\nonumber\\
&&
\left.
\hphantom{\left(\frac{\alpha_s}{\pi}\right)}
                  + (-0.01739- 0.2364\bar{l})\bar{\tau}_t^4
                  + (-0.03713 - 0.1928\bar{l})\bar{\tau}_t^5
\right)
\nonumber\\
&& 
         +\left(\frac{\alpha_s}{\pi}\right)^2
          \left(-1.292 - 1.75 \bar{l}
         +\left(-0.09881 - 0.7120\bar{l} 
             + 0.05833 {\bar{l}}\,^2 \right) \bar{\tau}_t\right.
\nonumber\\
&& 
\hphantom{\left(\frac{\alpha_s}{\pi}\right)}
          \left.
         +\left(0.2587 - 0.7879\bar{l} 
             + 0.4286 {\bar{l}}\,^2 \right) \bar{\tau}_t^2\right)
         +\ldots
         \Bigg]
\label{resms}
\end{eqnarray}
with $\bar{l}=\ln \mu^2/\bar{m}_t^2(\mu)$.
The leading order terms coincide in both schemes. This is the consequence
of the fact that in the limit of infinitely heavy top-quark mass $A_t$ does
not depend on $M_t$ at all.
In Fig.~\ref{fig3}, the results of ${\cal O}(\alpha_s^2)$ 
are plotted.
The curves show a similar behaviour as in the one- and two-loop cases.

From Eqs.~(\ref{resos}) and (\ref{resms}) the same observation can be made as
for earlier calculations up to ${\cal O}(\alpha_s^2)$
\cite{drho,KniSte95,CheKniSte96},
namely the coefficients of $\alpha_s/\pi$ are much smaller in the
$\overline{\mbox{MS}}$ scheme than in the on-shell scheme. Consider e.g.,
$\mu^2=M_t^2$ and $\mu^2=\bar{m}_t^2$ in Eqs.~(\ref{resos}) and (\ref{resms}).
Then, the ratio of the coefficients
in front of $(\alpha_s/\pi)^2\tau_t$ and $(\alpha_s/\pi)^2\tau_t^2$
is about 50 and 20, respectively. 
Another feature concerning the $\overline{\mbox{MS}}$ scheme
is already visible in the two-loop result of Fig.~\ref{fig12}. 
There the expansion in $1/M_t$ seems to converge more rapidly in the
$\overline{\mbox{MS}}$ than in the on-shell scheme. Actually,
the curves including corrections of order $\tau_t^3$, $\tau_t^4$
or $\tau_t^5$ are practically the same as the one where only
the ${\cal O}(\tau_t^2)$ corrections are present.
The dash-dotted curves in Fig.~\ref{fig3} indicate a similar behaviour.

To conclude, the ${\cal O}(\alpha_s^2)$ QCD corrections to the
decay process $H\to\gamma\gamma$ for an intermediate Higgs boson
were calculated.
We considered the contribution from a heavy top quark and evaluated the
first three terms in the expansion in the inverse top mass.
Inspired from the one- and two-loop case this should be
at least up to $M_H \approx 2M_W$
an excellent approximation to the exact result.
The leading term was confirmed using a low energy theorem and a method
based on the Fock-Schwinger gauge.

\vspace{2ex}                  
{\bf Acknowledgments}

\noindent
I would like to thank K.G. Chetyrkin, B.A. Kniehl and J.H. K\"uhn for very
interesting and useful discussions. I also thank A. Djouadi and
M. Spira for providing me with the data for the exact two-loop result.
I would also like to thank B.A. Kniehl for the invitation to 
the Ringberg Workshop and the generous hospitality.

\vspace{2ex}                  
{\bf References}


\begin{thebibliography}{99}

\bibitem{CDFD0} 
CDF Collaboration, F. Abe {\it et al.},
{\it Phys.\ Rev.\ Lett.\ }{\bf74} (1995) 2626;\\
D0 Collaboration, S. Abachi {\it et al.},
{\it Phys.\ Rev.\ Lett.\ }{\bf74} (1995) 2632.

\bibitem{Jan96}
P. Janot,
in {\it Proceedings of the Ringberg Workshop: The Higgs puzzle---What can we
learn from LEP~2, LHC, NLC, and FMC?}, Tegernsee, Germany, 8--13 December 1996,
edited by B.A. Kniehl (World Scientific, Singapore, to appear).

\bibitem{Ver91} 
J.A.M. Vermaseren, {\it Symbolic Manipulation with FORM},
(Computer Algebra Netherlands, Amsterdam, 1991).

\bibitem{EllGaiNan76}
J. Ellis, M.K. Gaillard and D.V. Nanopoulos, 
{\it Nucl. Phys.} {\bf B 106} (1976) 292. 

\bibitem{VaiVolSakShi79}
A.I. Vainshtein, M.B. Voloshin, V.I. Sakharov and M.A.Shifman,
{\it . J. Nucl. Phys.} {\bf 30} (1979) 711.

\bibitem{DjoSpiVdbZer91} 
A. Djouadi, M. Spira, J.J. van der Bij and P.M. Zerwas,
{\it Phys.\ Lett.\ } {\bf B 257} (1991)~187;\\
A. Djouadi, M. Spira and P.M. Zerwas,
{\it Phys. Lett.} {\bf B 311} (1993) 255.

\bibitem{DawKau93}
S. Dawson and R.P. Kauffman, {\it Phys.\ Rev.\ } {\bf D 47} (1993) 1264.

\bibitem{Stediss}
M. Steinhauser, Ph.D. Thesis, Univ. Karlsruhe (1996), unpublished.

\bibitem{KniSpi95} 
B.A. Kniehl and M. Spira, {\it Nucl.\ Phys.\ } {\bf B 443} (1995) 37; 
{\it Z. Phys.} {\bf C 69} (1995)~77.

\bibitem{KniSpi94} 
B.A. Kniehl and M. Spira, {\it Nucl.\ Phys.\ } {\bf B 432} (1994) 39.

\bibitem{KniSte95}
B.A. Kniehl and M. Steinhauser, {\it Phys. Lett.} {\bf B 365} (1996) 297;
{\it Nucl.\ Phys.\ }{\bf B 454} (1995) 485.

\bibitem{CheKniSte96}
K.G. Chetyrkin, B.A. Kniehl and M. Steinhauser, Report Nos. MPI/PhT/96-65
and hep-ph/9610456, to be published in {\it Phys. Rev. Lett.}.

\bibitem{Kil95} W. Kilian, {\it Z. Phys.} {\bf C 69} (1995) 89.

\bibitem{CheKueSte96}
  K.G. Chetyrkin, J.H. K\"uhn and M. Steinhauser,  
  {\it Phys.~Lett.} {\bf B 371} (1996) 93;
  MPI/Ph/96-27, TTP-96-13, hep-ph/9606230, 
  to be published in {\it Nucl.~Phys.} {\bf B}.

\bibitem{hphoton2}
B.A. Kniehl, K.G. Chetyrkin and M. Steinhauser, in preparation.

\bibitem{NovShiVaiZak84}
V.A. Novikov, M.A. Shifman, A.I. Vainshtein and V.I. Zakharov
{\it Fortschr. Phys.} {\bf 32} (1984) 585.

\bibitem{drho} 
L. Avdeev, J. Fleischer, S. Mikhailov and O. Tarasov,
{\it Phys.\ Lett.\ } {\bf B 336} (1994) 560; (E) {\bf B 349} (1995) 597;\\
K.G. Chetyrkin, J.H. K\"uhn and M. Steinhauser,
{\it Phys.\ Lett.\ } {\bf B 351} (1995) 331.

\bibitem{CheTka81Bro92} 
K.G. Chetyrkin and F.V. Tkachov, {\it Nucl.\ Phys.\ } {\bf B 192} (1981) 159;\\
D.J. Broadhurst, {\it Z. Phys.\ }{\bf C 54} (1992) 54.

\bibitem{GraBroGraSch90} 
N. Gray, D.J. Broadhurst, W. Grafe, and K. Schilcher,
{\it Z. Phys.\ }{\bf C 48} (1990) 673.

\end{thebibliography}
\end{document}